\documentclass[11pt]{article}
\usepackage[margin=0.95 in]{geometry}
\usepackage{amsmath}
\usepackage{amssymb,amsfonts}
\usepackage[all]{xy}
\usepackage{graphicx}
\usepackage[utf8x]{inputenc}
\usepackage{amsmath}
\usepackage{amssymb}
\usepackage{float}
\usepackage{array}
\usepackage{tikz}
\usepackage{mathtools}
\usepackage{mathrsfs}
\usepackage{hyperref}
\usepackage{cite}
\numberwithin{equation}{section}
\numberwithin{equation}{section}
\numberwithin{table}{section}

\newcommand{\be}{\begin{equation}}
\newcommand{\ee}{\end{equation}}

\begin{document}
\thispagestyle{empty}

\vspace*{3cm}
{}
\noindent
{\LARGE \bf  Anharmonic oscillators and the null bootstrap}
\vskip .4cm
\noindent
\linethickness{.06cm}
\line(10,0){467}
\vskip 1.1cm
\noindent
\noindent
{\large \bf Renjan Rajan John and Krishna Priya R}
\vskip 0.5cm
{\em 
\noindent
School of Pure and Applied Physics\\
  Mahatma Gandhi University, Kottayam, Kerala 686 560, India}
 \vskip 0.5cm
 {\em
 \noindent
renjan@mgu.ac.in
}
\vskip 1.2cm

\noindent {\sc Abstract: } We employ the technique of perturbative analytic null bootstrap to obtain the energy eigenvalues and ladder operators of the sextic anharmonic oscillator up to second order in the coupling. We confirm our results by deriving the same from traditional perturbation theory. We further perform the bootstrap approach on non-Hermitian PT symmetric Hamiltonians, focusing on the shifted harmonic oscillator and the celebrated cubic anharmonic oscillator.



\vskip1cm

\pagebreak

\newpage
\setcounter{tocdepth}{2}
\tableofcontents


\section{Introduction}
\label{sec-intro}
Following the enormous success enjoyed by the conformal bootstrap program (see \cite{Poland:2018epd} for a review), similar techniques were introduced to study  problems in quantum mechanics \cite{Lin:2020mme,Han:2020bkb}. 
Contrary to solving the Schr\"odinger equation by invoking a wave function, in the bootstrap approach one imposes positivity constraints and obtains bounds on the data that specify the theory. These data involve energy eigenvalues, couplings and expectation values of operators. However, not all of the expectation values are independent as they are related to each other by recursion relations. For each theory, the minimal set of independent data in terms which all the other expectation values can be expressed form the search space.

For a host of quantum mechanical problems the bootstrap program has successfully recovered previously known results and gone beyond. These involve studies on the hydrogen atom, harmonic and anharmonic oscillators, double well potentials and supersymmetric quantum mechanics among others \cite{Kazakov:2021lel,Berenstein:2021dyf,Bhattacharya:2021btd,Aikawa:2021eai,Berenstein:2021loy, Tchoumakov:2021mnh,Aikawa:2021qbl, Bai:2022yfv,Nakayama:2022ahr,Li:2022prn,Hu:2022keu,Blacker:2022szo,Berenstein:2022unr,Fan:2023bld}.
Most of the advances have been through numerical studies. Recently, an analytic perturbative bootstrap approach based on \cite{Li:2022prn} was introduced in \cite{Guo:2023gfi} and the quartic anharmonic oscillator was studied up to third order in the coupling. In this approach, null bootstrap equations were solved perturbatively in the quartic coupling to obtain expressions for the ladder operators and the energy eigenvalues. In the first part of our work, we make use of this technique and obtain the energy eigenspectrum and ladder operators for the sextic anharmonic oscillator. 

In the second part, we shift our focus from Hermitian theories to non-Hermitian PT symmetric theories \cite{Bender:1998ke,Dorey:2001uw,Mostafazadeh:2001jk,Mostafazadeh:2001nr,Mostafazadeh:2002id,Solombrino:2002vk,Bender:2002vv,Mostafazadeh:2002pd,Bender:2003fi,Mostafazadeh:2003gz,Bender:2004sa,Bender:2004vn,Mostafazadeh:2004qh,Mostafazadeh:2005wm,Bender:2005tb,Bender:2006xa,Bender:2007nj,Bender:2007wb,Mostafazadeh:2008pw,Mannheim:2017apd,Khan:2022uyz,Li:2022prn}. Interest in such theories was spurred   by the observation of  \cite{Bender:1998ke} that the eigenvalue spectrum of a class of such theories was real and positive. This was later rigorously proven in \cite{Dorey:2001uw}. To guarantee a quantum theory one further requires a positive definite norm and unitarity which ensures the conservation of probability. The conventional norm in which one considers inner product with the transpose complex conjugate fails to satisfy these conditions in non-Hermitian theories. Non-Hermitian PT symmetric theories were salvaged by the introduction of a variety of theory dependent modified norms which ensured that the requirements to define a consistent quantum theory were satisfied \cite{Mostafazadeh:2001jk,Mostafazadeh:2001nr,Mostafazadeh:2002id,Solombrino:2002vk,Bender:2004sa,Mostafazadeh:2004qh,Bender:2007wb,Mannheim:2017apd,Mostafazadeh:2005wm}. While non-hermiticity implies that conventional positivity constraints cannot be directly applied, with the help of an appropriately modified norm one can indeed have such constraints to implement the bootstrap program. The numerical bootstrap approach was performed on a large number of non-Hermitian PT symmetric quantum mechanical problems in \cite{Khan:2022uyz}. We analyse non-Hermitian PT symmetric anharmonic oscillators in the light of perturbative analytic bootstrap and obtain the eigenspectrum and ladder operators.

The plan for the rest of the paper is as follows. In Section~\ref{sec-notation}, we list the various notation that we employ in our paper. In Section~\ref{sec-rr}, we discuss and derive the recursion relations that relate various expectation values in a theory. In Section~\ref{sec-x6oscillator}, we apply the technique of analytic bootstrap to the sextic anharmonic oscillator. In Section~\ref{sec-PTNH}, we study non-Hermitian PT symmetric anharmonic oscillators.  In Appendix~\ref{app-TPTsection}, we derive the results for the sextic anharmonic oscillator from traditional perturbation theory and compare with the results obtained in Section~\ref{sec-x6oscillator}. 
%

\section{Notation}
\label{sec-notation}
In this section we list the essential notation that we use in this work. Most of these are standard, yet we list them for quick reference and ease of reading.
\begin{itemize}
\item $\mathbb Z_{\ge n}$ where $n$ is a non-negative integer denotes the set of integers greater than or equal to $n$.
\item $O$ denotes an arbitrary test operator that appears in the bootstrap equations. For our purpose it suffices to consider test operators formed out of products of position and momentum operators, i.e. 
\begin{align}
\label{test}
O=x^m(ip)^n
\end{align}
where $m\in \mathbb Z_{\ge 0}$ and $n\in \mathbb Z_{\ge 0}$. 
%
\item $|E\rangle$ denotes an eigenstate of the Hamiltonian $H$ with  eigenvalue $E$, i.e. we have the Schr\"odinger equation
\begin{align}
\label{eestate}
H|E\rangle=E|E\rangle
\end{align}
\item $g$, where $g\in\mathbb R$, denotes the coupling of the perturbative term in the Hamiltonian.
\item To avoid clutter, we will often denote the expectation value of an operator $O$ in the state $|E\rangle$, i.e.  $\langle E|O|E\rangle$, simply by  $\langle O\rangle_E$.
\item We call expectations of operators of the kind $O=x^m(ip)^n$, where $m$ and $n$ are both integers greater than zero, as mixed expectations.  Expectations of operators of the kind $O=x^t$, where $t\in\mathbb Z_{\ge 1}$, are called pure $x$ expectations. Expectations of operators of the kind $O=(ip)^t$, where $t\in\mathbb Z_{\ge 1}$, are called pure $p$ expectations.
\item We will use a perturbative expansion in $g$ of various quantities such as expectation values, energy eigenvalues, ladder operators etc. We use the following notation to denote the terms that appear in the expansion of a quantity, say $A$,
\begin{align}
\label{expnot}
A=\sum_{i=0}g^iA^{(i)}=A^{(0)}+gA^{(1)}+g^2A^{(2)}+\ldots
\end{align}
\item $L_{+}$ and $L_{-}$  denote the raising and lowering (ladder) operators.  
\item $|\text{ground}\rangle$ denotes the ground state of the theory.
\item  In non-Hermitian PT symmetric theories, we denote the suitably defined state that is conjugate to $|E\rangle$ by  $\langle\overline{E}|$.
\item We denote the Hermitian Hamiltonian that is equivalent to a non-Hermitian PT symmetric Hamiltonian $H$ by $\widetilde H$. We denote its eigenvalues by $\widetilde E$.
\end{itemize}
\section{Recursion relations}
\label{sec-rr}
In this section we rederive the basic recursion relations that we use in our study following \cite{Han:2020bkb}. Although in this work we focus on specific perturbations of the harmonic oscillator Hamiltonian, the recursion relations that we derive here apply to more general Hamiltonians of the kind
\begin{align}
\label{genH}
H=\frac{p^2}{2}+V(x)
\end{align}
The equations that we make use of follow from 
\begin{enumerate}
\item the definition of $|E\rangle$ \eqref{eestate}, and
\item the hermiticity of the Hamiltonian, which lets us write the following from \eqref{eestate}
\begin{equation}
\langle E|H=E\langle E|
\end{equation}
\end{enumerate}  %
From the above two points, one has the following 
\begin{align}
\label{fund1}
&\langle E|[H,O]|E\rangle=0\\[5pt]
\label{fund2}
&\langle E| HO|E\rangle=E\langle E| O|E\rangle
\end{align}
where $O$ is an arbitrary test operator \eqref{test}. In Section~\ref{sec-PTNH}, where we consider non-Hermitian PT symmetric Hamiltonians, the equations will be suitably modified.

We first consider the test operator $O=x^{t+1}$ where $t\in\mathbb Z_{\ge 0}$. We have from \eqref{fund1}
\begin{align}
\langle E|[H,x^{t+1}]|E\rangle=0
\end{align}
A straightforward application of the commutation relation $[p,x^{t}]=-itx^{t-1}$ gives
\begin{align}
\label{r1}
\langle x^{t}p\rangle=\frac i2t\langle x^{t-1}\rangle
\end{align}
where we dropped the subscript $E$ that denotes the energy eigenstate in which the expectation value is taken since \eqref{r1} is valid for all energy eigenstates.
Thus we have an expression for mixed expectations of the kind $\langle x^{t}p\rangle$ in terms of pure $x$ expectations $\langle x^{t-1}\rangle$. In particular, we have from setting $t=0$ in \eqref{r1}
\begin{align}
\label{r1p}
\langle p\rangle=0
\end{align}
Note that \eqref{r1} is independent of the specific theory since the potential $V(x)$ does not appear in it.

Let us now consider the test operator  $O=x^mp^{n-2}$, where $n\in\mathbb Z_{\ge 2}$. We get from \eqref{fund2}
\begin{align}
\label{r4}
\langle x^{m}p^n\rangle_E=2E\langle x^{m}p^{n-2}\rangle_E-2\langle x^{m}p^{n-2}V(x)\rangle_E
\end{align}
This gives us a recursion relation for mixed expectations of the kind $\langle x^{m}p^n\rangle_E$ in terms of mixed expectations which involve lower powers of $p$.  A special case of this recursion, which is often useful, is when $n=2$. This gives
\begin{align}
\label{r2}
\langle x^mp^2\rangle_E=2E\langle x^m\rangle_E-2\langle x^mV(x)\rangle_E
\end{align}
which gives an expression for mixed expectations of the kind $\langle x^{m}p^2\rangle$ in terms of pure $x$ expectations.
%
%

Let us now consider the test operator $O=x^tp$ in  \eqref{fund1}. A straightforward application of basic commutation relations and \eqref{r2}  leads to
\begin{align}
\label{r3}
t(t-1)(t-2)\langle x^{t-3}\rangle_E-8t\langle x^{t-1}V(x)\rangle_E+8tE\langle x^{t-1}\rangle_E-4\langle x^tV'(x)\rangle_E=0
\end{align}
This gives us an equation that relates various pure $x$ expectations.

We will now come to our analysis of the sextic anharmonic oscillator where we make use of the recursion relations derived here.

\section{Sextic anharmonic oscillator}
\label{sec-x6oscillator}
In this section we consider the following sextic anharmonic oscillator
%
\begin{align}
\label{anho}
H=\frac{p^2}{2}+\frac{x^2}{2}+gx^6
\end{align}
%
Since the theory is even under parity transformation we have
\begin{align}
\langle x^mp^n\rangle=0
\end{align}
whenever $m+n$ is an odd integer. In particular, we have 
\begin{align}
\langle x^m\rangle=\langle p^m\rangle=0,\quad\text{whenever}\,\,m\,\,\text{is\,\,odd}\,.
\end{align}
Recursion relations \eqref{r3} and \eqref{r4} take the following form
\begin{align}
\label{r3-x6}
t(t-1)(t-2)\langle x^{t-3}\rangle_E-4t\langle x^{t+1}\rangle_E-8gt\langle x^{t+5}\rangle_E+8tE\langle x^{t-1}\rangle_E-4\langle x^{t+1}\rangle_E-24g\langle x^{t+5}\rangle_E=0
\end{align}
\begin{align}
\label{r4-x6}
\langle x^{m}p^n\rangle_E=2E\langle x^{m}p^{n-2}\rangle_E-\langle x^{m}p^{n-2}x^2\rangle_E-2g\langle x^{m}p^{n-2}x^6\rangle_E
\end{align}
From \eqref{r3-x6} one can see that all pure $x$ expectations of the form $\langle x^t\rangle_E$ for $t\ge 6$ are given in terms of the expectations, $\langle x^2\rangle_E$ and $\langle x^4\rangle_E$, the energy $E$ and the coupling $g$. The first few expectations are given by
\begin{align}
\label{purexEg}
\langle x^6\rangle_E&=\frac{E-\langle x^2\rangle_E}{4g}\cr
\langle x^8\rangle_E&=\frac{3+12E\langle x^2\rangle_E-8\langle x^4\rangle_E}{24g}\cr
\langle x^{10}\rangle_E&=-\frac{3\left(E-\langle x^2\rangle_E\right)}{32g^2}+\frac{5\left(3\langle x^2\rangle_E+2E\langle x^4\rangle_E\right)}{16g}\cr
\langle x^{12}\rangle_E&=\frac{21\langle x^4\rangle_E}{8g}+\frac{-6+21E^2-45E\langle x^2\rangle_E+16\langle x^4\rangle_E}{120g^2}
\end{align}
We will now describe a procedure to express expectations in terms of energy at each order in perturbation.
\subsection{Expectations in terms of energy}
In this section, inspired by \cite{Guo:2023gfi}, we make two assumptions. 
\begin{enumerate}
\item The energy $E$, and the independent expectations, $\langle x^2\rangle_E$ and $\langle x^4\rangle_E$ in this case, have a perturbative expansion in $g$ as follows
\begin{align}
\label{seriesassumption}
E&=\sum_{i=0}g^iE^{(i)}=E^{(0)}+gE^{(1)}+g^2E^{(2)}+\ldots\cr
\langle x^2\rangle_E&=\sum_{i=0}g^i\langle x^2\rangle^{(i)}_E=\langle x^2\rangle^{(0)}_E+g\langle x^2\rangle_E^{(1)}+g^2\langle x^2\rangle_E^{(2)}+\ldots\cr
\langle x^4\rangle_E&=\sum_{i=0}g^i\langle x^4\rangle^{(i)}_E=\langle x^4\rangle^{(0)}_E+g\langle x^4\rangle_E^{(1)}+g^2\langle x^4\rangle_E^{(2)}+\ldots
\end{align}
\item Expectations do not blow up when the perturbation is turned off, i.e. in the $g\rightarrow 0$ limit expectations are non-singular and reduce to their expressions in the unperturbed theory.
\end{enumerate}
The second assumption helps fix the coefficients that appear in the expansion of $\langle x^2\rangle_E$ and $\langle x^4\rangle_E$ in \eqref{seriesassumption}, namely $\langle x^2\rangle^{(i)}_E$ and $\langle x^4\rangle^{(i)}_E$. Let us see this in detail. 

To obtain the zeroth order contribution to $\langle x^2\rangle_E$, i.e. $\langle x^2\rangle^{(0)}_E$, we consider the expectation $\langle x^6\rangle_E$ in \eqref{purexEg}. We make use of the expansion \eqref{seriesassumption} and impose the condition that $\langle x^6\rangle_E$ is non-singular in the $g\rightarrow 0$ limit. This gives 
\begin{align}
\label{x20}
\langle x^2\rangle^{(0)}_E=E^{(0)}
\end{align}
as expected from the unperturbed harmonic oscillator. To obtain the zeroth order contribution to $\langle x^4\rangle_E$, i.e. $\langle x^4\rangle^{(0)}_E$, we do the same analysis as above on $\langle x^8\rangle_E$ in \eqref{purexEg}. This gives
\begin{align}
\label{x40}
\langle x^4\rangle^{(0)}_E=\frac{3}{8}\left(1+4{E^{(0)}}^2\right)
\end{align}
Let us now extend this analysis to $\langle x^{10}\rangle_E$ in \eqref{purexEg}. The $1/g^2$  singularity is cured by \eqref{x20}. However, to cure the $1/g$ singularity, apart from \eqref{x20} and \eqref{x40}, we require the first order correction to $\langle x^2\rangle_E$, i.e. $\langle x^2\rangle^{(1)}_E$, to be as follows
\begin{align}
\langle x^2\rangle^{(1)}_E=-\frac{1}{2}\left(25E^{(0)}+20{E^{(0)}}^3-2E^{(1)}\right)
\end{align}
To fix the first order correction to $\langle x^4\rangle_E$, i.e. $\langle x^4\rangle^{(1)}_E$, we inspect $\langle x^{12}\rangle_E$ in \eqref{purexEg}. The $1/g^2$  singularity is cured by \eqref{x20} and \eqref{x40}. To get rid of the $1/g$ singularity, we  require the first order correction to $\langle x^4\rangle_E$, i.e. $\langle x^4\rangle^{(1)}_E$, to be as follows

\begin{align}
\langle x^4\rangle^{(1)}_E=-\frac{3}{128}\left(315+{2760E^{(0)}}^2+1200{E^{(0)}}^4-128E^{(0)}E^{(1)}\right)
\end{align}
This exercise can be performed to obtain the corrections to the independent expectations, $\langle x^2\rangle_E$ and $\langle x^4\rangle_E$, to any desired order in $g$.

Once we have obtained the coefficients $\langle x^2\rangle^{(i)}_E$ and $\langle x^4\rangle^{(i)}_E$, substituting them in the expressions for $\langle x^t\rangle_E$ for $t\ge 6$ (such as \eqref{purexEg}), obtained by solving \eqref{r3-x6}, gives us non-singular expressions for such expectations. Mixed expectations, $\langle x^{m}p^n\rangle_E$, are then obtained by the repeated application of \eqref{r4-x6} and finally \eqref{r1}. 
%

To determine pure $p$ expectations we make use of \eqref{r4-x6}  after setting $m=0$. This gives 
\begin{align}
\label{r4-x6-m=0}
\langle p^n\rangle_E=2E\langle p^{n-2}\rangle_E-\langle p^{n-2}x^2\rangle_E-2g\langle p^{n-2}x^6\rangle_E
\end{align}
Starting from $n=2$ we can obtain all the pure $p$ expectations from \eqref{r4-x6-m=0} by making use of the expressions for mixed expectations obtained in the previous step.
\subsection{Solving the sextic oscillator}
\label{sec-solvingsextic}
We will now apply the technique of null bootstrap to determine the energy spectrum and the ladder operators in the sextic theory.

The raising and lowering operators, $L_{+}$ and $L_{-}$ respectively, connect the energy eigenstates labelled by a non-negative integer $n$ as follows 
%
\begin{align}
\label{ladderaction1}
L_{+}|E_n\rangle&=\sqrt{n+1} |E_{n+1}\rangle\\
\label{ladderaction}
L_{-}|E_n\rangle&=\sqrt{n} |E_{n-1}\rangle
\end{align}
In the following we consider a perturbative expansion of the ladder operators in $g$
\begin{align}
\label{ladderseriesassumption}
L_{\pm}&=\sum_{i=0}g^iL_{\pm}^{(i)}=L_{\pm}^{(0)}+gL_{\pm}^{(1)}+g^2L_{\pm}^{(2)}+\ldots
\end{align}
We now describe the equations that we solve to obtain the $L_{\pm}^{(i)}$ in \eqref{ladderseriesassumption} and the $E_{n}^{(i)}$ in \eqref{seriesassumption}. \\ \\
\textbf{Schr\"odinger-like null state conditions}\\ 
These are of the form
%
\begin{align}
\label{SE1}
\langle E_n| O(H-E_{n+1})L_{+}|E_n\rangle&=0\\[5pt]
\label{SE2}
\langle E_n| O(H-E_{n-1})L_{-}|E_n\rangle&=0
\end{align}
where $O$ is an arbitrary test operator. These equations follow directly from the definition of the states $|E_{n}\rangle$  in \eqref{eestate} and from the action of the ladder operators \eqref{ladderaction}. \\ \\
\textbf{Null state condition}\\ 
We use the following null state condition to obtain the ground state energy at each order in $g$ 
\begin{align}
\label{NSC}
\langle\text{ground}|OL_{-}|\text{ground}\rangle=0
\end{align}
This follows directly from the definition of the ground state that it is annihilated by the lowering operator
\begin{align}
L_{-}|\text{ground}\rangle=0
\end{align}
Before we get into explicit calculations regarding solving \eqref{SE1}, \eqref{SE2} and \eqref{NSC}, we outline the procedure in the following sequence of steps \cite{Guo:2023gfi}. At each order in $g$, we do the following.
\begin{enumerate}
\item Make a judicious choice \footnote{A judicious choice points to a set that is not too restricted that we run into difficulties either while solving the equations or while imposing the remaining steps in the procedure and at the same time is not too large that it contains redundant data.} on the ansatz for the ladder operator and on the set of test operators. 
For ladder operators at $O(g^i)$ we consider the following ansatz
\begin{align}
\label{ansatzladder}
L^{(i)}_{\pm}=\sum_{m=0}^{K_i}\sum_{n=0}^{K_i-m}A^{(i)}_{m,n}x^m (ip)^n
\end{align}
where $K_i$ is a positive integer that is chosen at each order in $g$. Test operators are of the form in \eqref{test}.
\item Solve the set of equations, \eqref{SE1} for the raising operator (\eqref{SE2} for the lowering operator), one equation for each test operator, to obtain i) a recursion relation for the energy eigenvalues and ii) a subset of coefficients that appear in \eqref{ansatzladder} in terms of the remaining coefficients.
\item Use the null state condition, \eqref{NSC}, to obtain the ground state energy, and then use the recursion relation obtained in Step 2 to get the energy of higher levels.
\item Impose the condition that the ladder operator obtained in Step 2  is independent of energy. 
\item Impose the following normalisation condition
\begin{align}
\langle E_n|L_{+}^\dagger L_{+}|E_n\rangle&=n+1\\
\label{normalisation}
\langle E_n|L_{-}^\dagger L_{-}|E_n\rangle&=n
\end{align}
%
%
\end{enumerate}
%
%
%
Note that the results for the various expectations in the theory up to required orders in $g$ are absolutely crucial in Step 2 and Step 5.
In the following, we will apply the above sequence of steps order by order in $g$.
\subsubsection{$O(g^0)$}
\label{sec-o(1)sextic}
In this section we will obtain the ladder operator and the energy eigenvalues at $O(g^0)$, i.e. $L_{\pm}^{(0)}$ and $E_n^{(0)}$ by solving \eqref{SE1}, \eqref{SE2} and \eqref{NSC}. We will describe the computation for the lowering operator. Similar computations hold true for the raising operator and  we merely present the final results.

To solve \eqref{SE2} at $O(g^0)$, we set $K_i=1$ in  \eqref{ansatzladder}
\begin{align}
\label{ansatzg0}
L^{(0)}_{\pm}=\sum_{m=0}^1\sum_{n=0}^{1-m}A^{(0)}_{m,n}x^m (ip)^n
\end{align}
%
We substitute the ansatz in \eqref{SE2} and obtain a set of solutions for the coefficients $A^{(0)}_{m,n}$ and a recursion relation for the unperturbed energy eigenvalues. We pick the solution that corresponds to the lowering operator and obtain
\begin{align}
\label{energyrrg0}
E^{(0)}_{n-1}=E_{n}^{(0)}-1
\end{align}
as expected in the unperturbed theory. We also get the following conditions on the coefficients in \eqref{ansatzg0}
\begin{align}
\label{Astep1O1}
A^{(0)}_{0,0}=0,\quad A^{(0)}_{1,0}=A^{(0)}_{0, 1}
\end{align} 
Substituting \eqref{Astep1O1} in \eqref{ansatzg0}, we obtain
\begin{align}
\label{o1ladopint}
L_{-}^{(0)}&=A^{(0)}_{0,1}(x+ip)
\end{align}
We now apply the null state condition \eqref{NSC} and obtain the ground state energy
\begin{align}
\label{gstateenergyg0}
E_0^{(0)}=\frac 12
\end{align}		
as expected. We can now easily solve the recursion relation \eqref{energyrrg0} and obtain the following familiar result from the unperturbed oscillator
\begin{align}
E_n^{(0)}=n+\frac 12
\end{align}
We notice that the ladder operator obtained in \eqref{o1ladopint} is energy independent and hence we do not have to perform Step 4 from the sequence of steps given in Section \ref{sec-solvingsextic}. In the final step we perform a normalisation according to \eqref{normalisation} and obtain 
\begin{align}
\label{lowoptg0}
L_{-}^{(0)}=\frac{1}{\sqrt{2}}(x+ip)
\end{align} 
which matches the Dirac lowering operator for the unperturbed oscillator. 

A procedure along similar lines where after Step 1 we pick the solution that corresponds to the raising operator $E^{(0)}_{n+1}=E_{n}^{(0)}+1$  gives 
\begin{align}
\label{raisoptg0}
L_{+}^{(0)}=\frac{1}{\sqrt{2}}(x-ip)
\end{align} 
\subsubsection{$O(g)$}
\label{subsectionogx6}
Let us now obtain the correction to the ladder operators and energy eigenvalues at $O(g)$. To solve \eqref{SE2} at $O(g)$, we set $K_i=5$ in \eqref{ansatzladder}
\begin{align}
\label{ansatzg1}
L^{(1)}_{\pm}=\sum_{m=0}^5\sum_{n=0}^{5-m}A^{(1)}_{m,n}x^m (ip)^n
\end{align}
We substitute the ansatz in \eqref{SE2} and obtain the energy recursion relation
\begin{align}
\label{energrecg}
 E^{(1)}_{n-1} = \frac{1}{8} \left(-45 + 60 E_n^{(0)} - 60 {E_n^{(0)}}^2 + 8 E_n^{(1)}\right)
\end{align}
and the following relation between the coefficients in \eqref{ansatzg1}
\footnotesize
\begin{align}
\label{ogsolnlad}
 A^{(1)}_{0,0}&= 2 \left(-A^{(1)}_{0, 4} + A^{(1)}_{0, 2}E_n^{(0)}- 2 A^{(1)}_{0, 4} {E_n^{(0)}}^2\right)\cr
  A^{(1)}_{1, 0}&=
 \frac{1}{8} \left(15  \sqrt{2} + 8  A^{(1)}_{0,1} - 16 A^{(1)}_{0, 3} + 48 A^{(1)}_{0, 5} - 
     16 A^{(1)}_{1, 4} - 15 \sqrt{2} E_n^{(0)} - 16 A^{(1)}_{0, 3} E_n^{(0)} + 
     128 A^{(1)}_{0, 5}E_n^{(0)}\right. \cr
     &\hspace{1cm}\left.+ 16 A^{(1)}_{1, 2}E_n^{(0)}+ 30 \sqrt{2} {E_n^{(0)}}^2 + 
     32 A^{(1)}_{0, 5} {E_n^{(0)}}^2 - 32 A^{(1)}_{1, 4} {E_n^{(0)}}^2\right)\cr
     A^{(1)}_{1, 1}&= -2 \left(2 A^{(1)}_{0, 4} - A^{(1)}_{1, 3} E_n^{(0)} \right)\cr 
  A^{(1)}_{2, 0}&=-A^{(1)}_{0, 2} - 2 A^{(1)}_{1, 3} + 2 A^{(1)}_{0, 4} E_n^{(0)} - 2 A^{(1)}_{4, 0}E_n^{(0)}\cr
 A^{(1)}_{2, 1}&= 
 \frac{1}{16} \left(15 \sqrt{2} - 16 A^{(1)}_{0, 3} - 64 A^{(1)}_{1, 4} - 30 \sqrt{2} E_n^{(0)} + 
     64 A^{(1)}_{0, 5} E_n^{(0)}  + 32 A^{(1)}_{2, 3} E_n^{(0)}\right) \cr
    A^{(1)}_{2, 2}&= -A^{(1)}_{0, 4} - A^{(1)}_{4, 0},\cr 
A^{(1)}_{3, 0}&= 
 \frac {1}{16} \left(25\sqrt{2} - 128  A^{(1)}_{0, 5} - 16  A^{(1)}_{1, 2} - 32 A^{(1)}_{2, 3} - 
     10 \sqrt{2}  E_n^{(0)} + 64 A^{(1)}_{1, 4} E_n^{(0)}+ 32 A^{(1)}_{3, 2}E_n^{(0)}\right)\cr 
 A^{(1)}_{3, 1}&= - A^{(1)}_{1, 3}, \cr
  A^{(1)}_{4, 1}&= \frac{1}{8} \left(-5 \sqrt{2} - 8  A^{(1)}_{0, 5} - 8  A^{(1)}_{2, 3}\right) \cr
  A^{(1)}_{5, 0}&= \frac{1}{8} \left(-\sqrt{2} - 8  A^{(1)}_{1, 4} - 8  A^{(1)}_{3, 2}\right)
\end{align}
\normalsize
Substituting \eqref{ogsolnlad} in \eqref{ansatzg1}, we obtain
\footnotesize	
\begin{align}
\label{ladogint}
L_{-}^{(1)}&=i p A^{(1)}_{0, 1} - 2 A^{(1)}_{0, 4}+ 
 x \left(\frac{15}{4 \sqrt{2}} + A^{(1)}_{0, 1}- 2 A^{(1)}_{0, 3} + 6 A^{(1)}_{0, 5} - 2 A^{(1)}_{1, 4}\right) - 
 A^{(1)}_{0, 2} p^2 - 4 i A^{(1)}_{0, 4} x p \cr
 &\hspace{0.2cm}+ \left(-A^{(1)}_{0, 2} - 2 A^{(1)}_{1, 3}\right) x^2 - 
 i A^{(1)}_{0, 3} p^3 - A^{(1)}_{1, 2} x p^2 + 
 \frac{i}{16 } \left(15 \sqrt{2} - 16 A^{(1)}_{0, 3} - 64 A^{(1)}_{1, 4}\right) x^2 p\cr 
 &\hspace{0.2cm}+ \left(\frac{25}{
    8 \sqrt{2}} - 8 A^{(1)}_{0, 5} - A^{(1)}_{1,2}- 2 A^{(1)}_{2, 3}\right) x^3 + A^{(1)}_{0, 4}p^4 - 
 i A^{(1)}_{1, 3} x p^3 +\left (A^{(1)}_{0, 4} + A^{(1)}_{4, 0}\right) x^2 p^2\cr
  &\hspace{0.2cm}- i A^{(1)}_{1, 3}x^3 p + 
 A^{(1)}_{4, 0} x^4 + i A^{(1)}_{0, 5} p^5 + A^{(1)}_{1, 4}xp^4 - i A^{(1)}_{2, 3} x^2 p^3 - 
 A^{(1)}_{3, 2} x^3 p^2 \cr
 &\hspace{0.2cm}-\frac{1}{8} i \left(5 \sqrt{2} + 8 A^{(1)}_{0, 5} + 8A^{(1)}_{2, 3}\right) x^4 p + \left(\frac{-1}{4 \sqrt{2}} -
     A^{(1)}_{1, 4} - A^{(1)}_{3, 2}\right) x^5 \cr
     &\hspace{0.2cm}+ \left[2 A^{(1)}_{0, 2} + 
    x \left(\frac{-15}{4 \sqrt{2}} - 2 A^{(1)}_{0, 3}+ 16 A^{(1)}_{0, 5} + 2 A^{(1)}_{1, 2}\right) + 
    2 i A^{(1)}_{1, 3} x p + 
    2 (A^{(1)}_{0, 4}- A^{(1)}_{4, 0}) x^2 \right.\cr
    &\hspace{0.2cm}+\left.\left (-\frac{15 i}{4 \sqrt{2}} + 
       4 i A^{(1)}_{0, 5} + 2 i A^{(1)}_{2, 3}\right) x^2 p + \left(\frac{-5}{4 \sqrt{2}} + 
       4 A^{(1)}_{1, 4}+ 2 A^{(1)}_{3, 2}\right) x^3\right]E_n^{(0)}\cr
   &\hspace{0.2cm}+\left[-4 A^{(1)}_{0, 4} + x \left(\frac{15}{2 \sqrt{2}} + 4 A^{(1)}_{0, 5} - 4 A^{(1)}_{1, 4}\right)\right]{E_n^{(0)}}^2
\end{align}
\normalsize
To obtain the first order correction to the ground state energy, we apply the null state condition \eqref{NSC} and obtain 
\begin{align}
E_0^{(1)}=\frac{15}{8}
\end{align}
Combined with the energy recursion relation \eqref{energrecg}, we obtain the first order correction  
\begin{align}
\label{En1B}
E_n^{(1)}=\frac{15}{8}+5n+ \frac{15 n^{2}}{4} + \frac{5 n^{3}}{2}
\end{align}
This matches the result obtained in \eqref{en1app} from traditional perturbation theory.

The next step involves ensuring that the ladder operator \eqref{ladogint} is energy independent. Imposing this we get
\footnotesize
\begin{align}
A^{(1)}_{0, 2} &= 0,\, A^{(1)}_{1, 3}=0,\, A^{(1)}_{4, 0}=0,\, A^{(1)}_{0, 4}=0\cr 
 A^{(1)}_{2, 3} &=\frac{1}{16} \left(15 \sqrt{2} - 32 A^{(1)}_{0, 5}\right)  \cr
 A^{(1)}_{3, 2} &= \frac{1}{16} \left(-25 \sqrt{2} - 32 A^{(1)}_{0, 5}\right)  \cr
 A^{(1)}_{1, 2} &= \frac{15}{8\sqrt{2}} +  A^{(1)}_{0, 3}-8A^{(1)}_{0, 5} \cr
 A^{(1)}_{1, 4} &= \frac{1}{16} \left(15 \sqrt{2} + 16 A^{(1)}_{0, 5}\right) 
\end{align}
\normalsize
Substituting in \eqref{ladogint} we get
\footnotesize
    \begin{align}
    \label{laddopgpen}
L_{-}^{(1)}&=ip A^{(1)}_{0, 1} + x \left(A^{(1)}_{0, 1} - 2 A^{(1)}_{0, 3} + 4 A^{(1)}_{0, 5}\right) - 
 iA^{(1)}_{0, 3} p^3 + \left(-\frac{15}{8 \sqrt{2}} - A^{(1)}_{0, 3} + 8 A^{(1)}_{0, 5}\right) xp^2\cr 
 &- \frac{i}{16}  \left(45 \sqrt{2} + 16 A^{(1)}_{0, 3} + 
    64A^{(1)}_{0, 5}\right) x^{2}p + \left(\frac{-5}{2 \sqrt{2}} - A^{(1)}_{0, 3} + 
    4 A^{(1)}_{0, 5}\right) x^3 \cr
    &+ \frac{1}{16} \left(16 i A^{(1)}_{0, 5} p^5 + \left(15 \sqrt{2} + 16 A^{(1)}_{0, 5}\right) xp^4 - 
    i \left(15 \sqrt{2} - 32 A^{(1)}_{0, 5}\right) x^2p^3 + \left(25 \sqrt{2} + 
       32 A^{(1)}_{0, 5}\right) x^3p^2\right) \cr
       &+ \left(-\frac{25 i}{8 \sqrt{2}} + 
    iA^{(1)}_{0, 5}\right) x^4p + \left(\frac{1}{\sqrt{2}} + A^{(1)}_{0, 5}\right) x^5
    \end{align}
    \normalsize
We then make use of the normalisation condition \eqref{normalisation} and get
\footnotesize
\begin{align}
A^{(1)}_{0,5}=-\frac{5}{16\sqrt{2}},\quad A^{(1)}_{0,3}=\frac{25}{8\sqrt{2}},\quad A^{(1)}_{0,1}=\frac{15}{4\sqrt{2}}
\end{align} 
\normalsize
Substituting for these coefficients in \eqref{laddopgpen}, we obtain the first order correction to the lowering operator
\begin{align}
\label{lowoptg1}
L_{-}^{(1)}&=\frac{15ip}{4\sqrt{2}}-\frac{15x}{4\sqrt{2}}-\frac{25ip^3}{8\sqrt{2}}-\frac{15xp^2}{2\sqrt{2}}-\frac{15ix^2p}{2\sqrt{2}}-\frac{55x^3}{8\sqrt{2}}-\frac{5ip^5}{16\sqrt{2}}+\frac{25xp^4}{16\sqrt{2}}-\frac{5ix^2p^3}{2\sqrt{2}}\nonumber\\[5pt]
&\hspace{0.3cm}+\frac{5x^3p^2}{2\sqrt{2}}-\frac{55ix^4p}{16\sqrt{2}}+\frac{11x^5}{16\sqrt{2}}
\end{align}
One can perform a similar analysis and obtain the first order correction to the raising operator
\begin{align}
\label{raisoptg1}
L_{+}^{(1)}&=-\frac{15ip}{4\sqrt{2}}-\frac{15x}{4\sqrt{2}}-\frac{25ip^3}{8\sqrt{2}}+\frac{15xp^2}{2\sqrt{2}}-\frac{15ix^2p}{2\sqrt{2}}+\frac{55x^3}{8\sqrt{2}}+\frac{5ip^5}{16\sqrt{2}}+\frac{25xp^4}{16\sqrt{2}}+\frac{5ix^2p^3}{2\sqrt{2}}\nonumber\\[5pt]
&\hspace{0.5cm}+\frac{5x^3p^2}{2\sqrt{2}}+\frac{55ix^4p}{16\sqrt{2}}+\frac{11x^5}{16\sqrt{2}}
\end{align}
We can check that these operators are Hermitian conjugates of each other, i.e.
\begin{align}
L_{+}^{(1)}&={L_{-}^{(1)}}^\dagger
\end{align}
They also match the expressions for the same obtained in \eqref{laddappexp} and \eqref{laddappexp1} using traditional perturbation theory, when the latter are expressed in terms of $x$ and $p$ operators using \eqref{App-aadaggerdef}.
\subsubsection{$O(g^2)$}
We will now move to the next order in the coupling. To solve \eqref{SE2} at $O(g^2)$, we set $K_i=9$ in \eqref{ansatzladder}
\begin{align}
\label{ansatzg2}
L^{(2)}_{\pm}=\sum_{m=0}^9\sum_{n=0}^{9-m}A^{(2)}_{m,n}x^m (ip)^n
\end{align}
Proceeding exactly as in Section \ref{subsectionogx6}, we obtain the following for $E_{n}^{(2)}$ and $L_{\pm}^{(2)}$
%
\begin{align}
\label{En2B}
E_{n}^{(2)}=\frac{1}{64}\left(-3495- 11528 n- 14400 n^2- 12220 n^3-3930 n^4- 1572 n^5\right)
\end{align}
and
%
\begin{align}
\label{lowoptg2}
L_{\pm}^{(2)}&=\mp\frac{15135 i p}{32 \sqrt{2}} - \frac{17985 x}{32\sqrt{2}} - \frac{
 167035 i p^3 }{256\sqrt{2}}\pm \frac{194085xp^2}{128 \sqrt{2}}  
  - \frac{237435 i x ^2p}{128 \sqrt{2}} \pm \frac{
 236085 x^3 }{256 \sqrt{2}}\nonumber\\[5pt]
  &\hspace{0.5cm}\pm \frac{86621 i p^5 }{
 512 \sqrt{2}}
+ \frac{439105xp^4 }{512 \sqrt{2}} \pm \frac{
 169285 i x^2p^3}{128 \sqrt{2}} + \frac{
 206235 x^3 p^2}{128\sqrt{2}} \pm \frac{
 608655 ix^4 p}{512 \sqrt{2}} + \frac{
 124371x^5}{512\sqrt{2}} \nonumber\\[5pt]
&\hspace{0.5cm}  + \frac{2381 i p^7}{64\sqrt{2}}
\mp \frac{25207 x p^6}{256 \sqrt{2}} + \frac{
 81621 ix^2 p^5}{256 \sqrt{2}} \mp\frac{
 52295 x^3p^4}{128 \sqrt{2}} + \frac{
 63545 ix^4p^3}{128 \sqrt{2}} \mp \frac{
 111771 x^5p^2}{256 \sqrt{2}}\nonumber\\[5pt]
 &\hspace{0.5cm}+ \frac{41657 ix^6p}{256 \sqrt{2}}
 \mp \frac{4771 x^7}{64 \sqrt{2}} \mp \frac{
 167 i p^9}{128 \sqrt{2}} - \frac{
 2381 xp^8}{256 \sqrt{2}} \mp \frac{
 3601 i x^2 p^7}{256 \sqrt{2}} - \frac{
 9069 x^3p^6}{256 \sqrt{2}} \mp \frac{
 10459 i x^4 p^5}{256 \sqrt{2}}\nonumber\\[5pt]
 &\hspace{0.5cm}- \frac{12709x^5 p^4}{256 \sqrt{2}}\mp \frac{
 12419 i x^6 p^3}{256 \sqrt{2}} - \frac{
 5951 x^7 p^2}{256 \sqrt{2}} \mp\frac{
 4771 i x^8 p}{256 \sqrt{2}} - \frac{
 97 x^9}{128 \sqrt{2}}
 \end{align}
%
%
These precisely match the results from traditional perturbation theory obtained in \eqref{en2app}, \eqref{laddappexp} and \eqref{laddappexp1}. As one would like it, $L_{\pm}^{(2)}$ operators satisfy
\begin{align}
L_{+}^{(2)}&={L_{-}^{(2)}}^\dagger
\end{align}
The above bootstrap approach can be extended to any desired order in $g$.
%
%
\section{PT symmetric non-Hermitian theory}
\label{sec-PTNH}
In this section, we consider PT symmetric non-Hermitian theories. The action of parity (P) and time reversal (T) transformations are as follows
\begin{align}
\label{PTaction}
\text{P} &: x\rightarrow -x,\quad p\rightarrow -p\cr
\text{T} &: x\rightarrow x,\quad p\rightarrow -p,\quad i\rightarrow -i
\end{align}
As mentioned in the Introduction, PT symmetric non-Hermitian theories must have a modified inner product that helps define a quantum theory.
This requires introducing an operator $S$ such that \cite{Mostafazadeh:2001jk,Mostafazadeh:2001nr,Mostafazadeh:2002id,Solombrino:2002vk,Bender:2004sa,Mostafazadeh:2004qh,Mannheim:2017apd}
\begin{align}
\label{HHdef}
SHS^{-1}=\widetilde H
\end{align}
where $\widetilde H$ is Hermitian. The orthonormality of the eigenstates of $\widetilde H$ ensures that the eigenstates of the original Hamiltonian $H$ satisfy the following modified orthonormality condition \cite{Mostafazadeh:2001jk,Mostafazadeh:2001nr,Mostafazadeh:2002id,Solombrino:2002vk,Bender:2004sa,Mostafazadeh:2004qh,Mannheim:2017apd}
\begin{align}
\label{vnormON}
    \langle \overline{E_m}|E_n\rangle=\delta_{m,n}
\end{align}
where $\langle \overline{E_m}|$ denotes the modified conjugate state to $|E_n\rangle$ defined as $\langle\overline{E_m}|=\langle E_m|V$ where $V$ is the positive Hermitian operator given by 
\begin{align}
V=S^\dagger S
\end{align}
The hermiticity of $\widetilde H$ implies that $V$ has the following action on $H$
\begin{align}
\label{Vdef}
VHV^{-1}=H^\dagger
\end{align}
%

%
%
Let us now see how the discussions in Section \ref{sec-rr} and Section \ref{sec-x6oscillator} are modified in non-Hermitian PT symmetric theories. Equation \eqref{fund1} is adapted to this case in the following form 
\begin{equation}
\label{fund1nh}
\langle\overline{E}|[H,O]|E\rangle=0
\end{equation} 
This can be shown to hold as follows
\begin{align}
\langle\overline{E}|[H,O]|E\rangle
&=\langle E|V(HO-OH)|E\rangle\cr
&=\langle E|VHV^{-1}VO-VOH|E\rangle\cr
&=\langle E|(H^\dagger VO-VOH)|E\rangle\cr
&=\langle E| E^\star VO-VOE|E\rangle\cr
&=(E^\star-E)\langle E| VO|E\rangle\cr
&=0
\end{align}
where the first line of the R.H.S follow from the definition of the conjugate state below \eqref{vnormON}, the third line from \eqref{Vdef} and the last line from the fact that the energy eigenspectrum is real.

Equation \eqref{fund2} in a non-Hermitian PT symmetric theory takes the form 
\begin{equation}
\label{fund2nh}
\langle\overline{E}|OH|E\rangle=E\langle\overline{E}|O|E\rangle
\end{equation} 
The Schr\"odinger-like null state conditions \eqref{SE1}, \eqref{SE2} and the null state condition \eqref{NSC} written using the modified norm take the form below
\begin{align}
\label{SE1nh}
\langle\overline{E_n}| O(H-E_{n+1})L_{+}|E_n\rangle=0
\end{align}
\begin{align}
\label{SE2nh}
\langle\overline{E_n}| O(H-E_{n-1})L_{-}|E_n\rangle=0
\end{align}
\begin{align}
\label{NSCnh}
\langle\overline{\text{ground}}|OL_{-}|\text{ground}\rangle=0
\end{align}
We will now  use these equations to perform the null bootstrap of non-Hermitian, PT symmetric theories. The procedure follows that of the computation of the sextic oscillator in Section \ref{sec-x6oscillator}. The recursion relations derived in Section \ref{sec-rr} will continue to hold, once we have appropriately modified the inner products as discussed above.

\subsection{Shifted harmonic oscillator}
\label{subsec-shifted}
Consider a shifted harmonic oscillator with the following Hamiltonian \cite{Bender:2007wb,Khan:2022uyz}
\begin{align}
\label{shifted}
H=\frac{p^2}{2}+\frac{x^2}{2}+igx
\end{align}
Although the Hamiltonian is not Hermitian, its PT invariance is evident from  \eqref{PTaction}.

The operator  $V$ \eqref{Vdef} that defines the appropriate conjugate state in this theory is given by \cite{Bender:2007wb,Khan:2022uyz}
\begin{align}
V=e^{2gp}
\end{align}
One can easily check using the Baker-Campbell-Hausdorff (BCH) formula that \eqref{Vdef} holds exactly
\begin{align}
e^{2gp}\left(\frac{p^2}{2}+\frac{x^2}{2}+igx\right)e^{-2gp}=\left(\frac{p^2}{2}+\frac{x^2}{2}-igx\right)
\end{align}
It is crucial that the $O(g^2)$ terms arising from a simple commutator of $e^{2gp}$ and $igx$ in the BCH formula cancel the ones that arise from nested commutators of $e^{2gp}$ and $\frac{p^2}{2}+\frac{x^2}{2}$. The equivalent Hermitian Hamiltonian \eqref{HHdef} is given by
\begin{align}
\widetilde H=e^{gp}\left(\frac{p^2}{2}+\frac{x^2}{2}+igx\right)e^{-gp}=\left(\frac{p^2}{2}+\frac{x^2}{2}+\frac{g^2}{2}\right)
\end{align}
The energy spectrum of such a Hamiltonian is straightforwardly given by a constant $O(g^2)$ shift of the simple harmonic oscillator spectrum 
\begin{align}
\label{shifted energy exp}
E_n=\left(n+\frac 12\right)+\frac{g^2}{2}
\end{align}
Let us now apply the technique of null bootstrap and obtain the corrections to the ladder operators and the eigenspectrum. 
%
We first consider the test operator $O=p^t$ in \eqref{fund1nh}
\begin{align}
\label{fund1nhO1}
\langle\overline{E}|[H,p^t]|E\rangle=0
\end{align}
Using basic commutation relations such as $[x^2,p^t]=2itp^{t-1}x-t(t-1)p^{t-2}$ we get from \eqref{fund1nhO1}
\begin{align}
\label{r1nh}
\langle\overline{E}|p^{t-1}x|E\rangle=-ig\langle\overline{E}|p^{t-1}|E\rangle-\frac i2(t-1)\langle \overline{E}|p^{t-2}|E\rangle
\end{align}
For $t=1$, this directly gives 
\begin{align}
\label{xexp}
\langle\overline{E}|x|E\rangle&=-ig\langle\overline{E}|E\rangle\cr
&=-ig
\end{align}
where we used orthonormality as in \eqref{vnormON}.

Let us now use the same test operator, $O=p^t$, in \eqref{fund2nh}. This gives 
\begin{align}
\label{r2nh}
\langle\overline{E}|p^tx^2|E\rangle&=2(E-g^2)\langle\overline{E}|p^t|E\rangle-gt\langle\overline{E}|p^{t-1}|E\rangle-\langle\overline{E}|p^{t+2}|E\rangle
\end{align}
In \eqref{r1nh} and \eqref{r2nh} we have obtained recursion relations for mixed expectations of the kinds $\langle\overline{E}|p^tx|E\rangle$ and $\langle\overline{E}|p^tx^2|E\rangle$ in terms of pure $p$ expectations.

Let us now consider \eqref{fund1nh} with the test operator $O=p^tx$, i.e.
\begin{align}
\langle\overline{E}|[H,p^tx]|E\rangle=0
\end{align}
Making use of \eqref{r1nh} and \eqref{r2nh} this leads to an equation that relates pure $p$ expectations
\begin{align}
\label{pureprelnh}
8tE \langle\overline{E}|p^{t - 1}|E\rangle- 4t g^2 \langle\overline{E}|p^{t - 1}|E\rangle+t(t - 1)(t - 2)\langle\overline{E}|p^{t - 3}|E\rangle- 
 4(t + 1) \langle\overline{E}|p^{t + 1}|E\rangle=0
 \end{align}
 We first note that setting $t=0$ in \eqref{pureprelnh}
 \begin{align}
  \langle\overline{E}|p|E\rangle=0
 \end{align}
Larger positive even values for $t$ in \eqref{pureprelnh} gives 
 \begin{align}
  \langle\overline{E}|p^m|E\rangle=0,\quad\text{whenever}\,\,m\,\,\text{is\,\,odd}\,.
 \end{align}
We also see that all the non-zero pure $p$ expectations are expressed in terms of the energy $E$ and the coupling $g$. We list down the first few of them explicitly
 \begin{align}
 \label{pexpappig}
 \langle\overline{E}|p^2|E\rangle &=E -\frac{g^{2}}{2}  \cr
 \langle\overline{E}|p^4|E\rangle &= \frac{3}{8} (1+4 E^2)-\frac {3}{2} g^2 E +\frac{3 g^4}{8} \cr
 \langle\overline{E}|p^6|E\rangle &= \frac{5}{8}E(5+4 E^2)-\frac {5}{16} g^2 (5+12E^2) +\frac{15E g^4}{8}-\frac{5g^6}{16}
  \end{align}
We will also find it useful to consider $O=p^mx^{n-2}$, where $m\in\mathbb Z_{\ge 0}$ and $n\in\mathbb Z_{\ge 2}$, in \eqref{fund2nh}. This gives
 \begin{align}
 \label{eq(21)}
\langle\overline{E}|p^{m}x^{n}|E\rangle= 
 2E\langle\overline{E}|p^{m}x^{n - 2}|E\rangle - \langle\overline{E}|p^{m}x^{n - 2}p^2|E\rangle - 2 ig\langle\overline{E}|p^{m }x^{n - 1}|E\rangle
 \end{align}
 In particular, for $m=0$ we get
  \begin{align}
 \label{eq(22)}
\langle\overline{E}|x^{n}|E\rangle= 
 2E\langle\overline{E}|x^{n - 2}|E\rangle - \langle\overline{E}|x^{n - 2}p^2|E\rangle - 2 ig\langle  \overline{E}|x^{n - 1}|E\rangle
 \end{align}
While we have already obtained $\langle\overline{E}|x|E\rangle$ in \eqref{xexp}, equations \eqref{eq(22)} and \eqref{eq(21)} help us obtain higher pure $x$ expectations in terms of pure $p$ expectations. After making use of the pure $p$ expectations obtained from \eqref{pureprelnh}, we will be able to express all the higher pure $x$ expectations in terms of the energy $E$ and the coupling $g$.
Some of the first few such expectations  are as follows
\begin{align}
\langle\overline{E}|x^{2}|E\rangle&= E- \frac 32 g^2 \cr
\langle\overline{E}|x^{3}|E\rangle&= -3 i g E
\end{align}
To  obtain  mixed expectations, we make use of \eqref{eq(21)} recursively and the pure $p$ expectations obtained earlier. 
Once we have the relations between the various expectations, we are ready to perform the null bootstrap. We proceed exactly as in the Hermitian sextic case. At $O(g^0)$, the theory \eqref{shifted} reduces to the simple harmonic oscillator, and the results for the ladder operators and spectrum match that of Section \ref{sec-o(1)sextic} and we do not repeat it here.
\subsubsection{$O(g)$}
We will now set out to solve \eqref{SE2nh} at $O(g)$. As before we focus on the lowering operator. We set $K_i=3$ in the ansatz for the lowering operator \eqref{ansatzladder}
\begin{align}
\label{ansatzg1nh}
L^{(1)}_{-}=\sum_{m=0}^3\sum_{n=0}^{3-m}A^{(1)}_{m,n}x^m (ip)^n
\end{align}
Proceeding exactly as in Section \ref{subsectionogx6} we obtain the following energy recursion relation
\begin{align}
E_n^{(1)}=E^{(0)}_{n-1}
\end{align}
The null state condition \eqref{NSCnh} gives the correction to the ground state energy to be
\begin{align}
E_0^{(1)}=0
\end{align}
Combined with the recursion relation obtained above we get
\begin{align}
E_n^{(1)}=0
\end{align} 
For the ladder operator, after solving \eqref{SE2nh} and imposing its energy independence and normalisation as below
\begin{align}
\label{normalisationnh}
\langle\overline{E_n}|L_{-}^\dagger L_{-}|E_n\rangle&=n
\end{align}
we get 
\begin{align}
L_{-}^{(1)}&=\frac{i}{\sqrt{2}}
\end{align}
Similarly, one obtains for the raising operator 
\begin{align}
L_{+}^{(1)}&=\frac{i}{\sqrt{2}}
\end{align}
Let us now consider the $O(g^2)$ corrections.
\subsubsection{$O(g^2)$}
Solving  \eqref{SE2nh} at $O(g^2)$ gives the following energy recursion relation 
\begin{align}
E_n^{(2)}=E_{n-1}^{(2)}
\end{align}
Applying the null state condition \eqref{NSCnh} at $O(g^2)$ gives 
\begin{align}
E_0^{(2)}=\frac 12
\end{align}
Combined with the recursion relation obtained above we get 
\begin{align}
E_n^{(2)}=\frac 12
\end{align}
This matches the energy spectrum of the equivalent Hamiltonian $\widetilde H$ derived in \eqref{shifted energy exp}. 

We also see that the corrections to the ladder operators vanish at $O(g^2)$, i.e. 
\begin{align}
L_{\pm}^{(2)}=0
\end{align}
Thus the Hamiltonian \eqref{shifted} is expressed in terms of  ladder operators as 
\begin{align}
H=L_{+}L_{-}+\frac 12+\frac{g^2}{2}
\end{align}
as in the case of the simple harmonic oscillator.
%

\subsection{Cubic anharmonic oscillator}
In this section we consider the non-Hermitian PT symmetric cubic anharmonic oscillator 
%
\begin{align}
\label{igx3}
H=\frac{p^2}{2}+\frac{x^2}{2}+igx^3
\end{align}
This Hamiltonian was shown to have a real and positive eigenspectrum in \cite{Bender:1998ke,Dorey:2001uw}. 
The operator $V$ can be written as an exponential \cite{Mostafazadeh:2001jk,Mostafazadeh:2001nr,Mostafazadeh:2002id,Bender:2004sa,Mostafazadeh:2004qh}
\begin{align}
\label{Qdef}
V=e^{-Q}
\end{align}
where $Q$ was derived in a perturbative series in $g$ and was shown to take the following form \cite{Bender:2004sa,Mostafazadeh:2004qh}
\begin{align}
Q&=\sum_{i=2k+1, k\in\mathbb Z_{\ge 0}}g^k\,Q^{(k)}\cr
&=g\,Q^{(1)}+g^3\,Q^{(3)}+\ldots
\end{align}
where the first few $Q^{(i)}$ are given by
\begin{align}
Q^{(1)}&=-\left(\frac 43p^3+2xpx\right)\cr
Q^{(3)}&=\frac{128}{15}p^5+\frac{40}{3}xp^3x+8x^2px^2-32p
\end{align}
One can check that the $V$ operator in \eqref{Qdef} satisfies
\begin{align}
\label{Qactionigx3}
e^{-Q}\left(\frac{p^2}{2}+\frac{x^2}{2}+igx^3\right)e^{Q}=\left(\frac{p^2}{2}+\frac{x^2}{2}-igx^3\right)
\end{align}
as required by \eqref{Vdef}.
The equivalent Hermitian Hamiltonian \eqref{HHdef} in this case is given by 
\begin{align}
\label{eqHigx3}
\widetilde H&=e^{-Q/2}\left(\frac{p^2}{2}+\frac{x^2}{2}+igx^3\right)e^{Q/2}\cr
&=\frac{p^2}{2}+\frac{x^2}{2}+\frac{g^2}{2}\left(-4-12ixp+6x^2p^2+3x^4\right)
\end{align}
where the last term on the R.H.S can be checked to be Hermitian. Once we have the equivalent Hermitian Hamiltonian we can use  traditional perturbation theory to find the corrections to the spectrum. Notice that \eqref{eqHigx3} has its first perturbation appearing at $O(g^2)$ and hence one can employ first order perturbation theory to compute the energy corrections. In particular, one has at  $O(g^2)$
\begin{align}
\label{eqHev}
\widetilde E^{(1)}_n&=\prescript{(0)}{}\langle E_{n}|\frac 12\left(-4-12ixp+6x^2p^2+3x^4\right)|E_n\rangle^{(0)}\cr
&=\frac 18\left(30n^2+30n+11\right)
\end{align}
%
Let us now apply null bootstrap techniques to the non-Hermitian PT symmetric Hamiltonian in \eqref{igx3} and obtain the corrections to the ladder operators and the eigenspectrum. The recursion relation derived in \eqref{r3} and \eqref{r4} take the following form
\begin{align}
\label{rr3}
&t(t-1)(t-2)\langle \overline{E}|x^{t-3}|E\rangle-4t\langle\overline{E}| x^{t+1}|E\rangle+8tE\langle\overline{E}| x^{t-1}|E\rangle_E-4\langle\overline{E}| x^{t+1}|E\rangle\nonumber\\[5pt]
&\hspace{0.25cm}-12 ig\langle \overline{E}|x^{t+2}|E\rangle-8itg\langle\overline{E}| x^{t+2}|E\rangle=0\\[5pt]
\label{rr32}
&\langle \overline{E}| x^{m}p^n|E\rangle=2E\langle \overline{E}|x^{m}p^{n-2}|E\rangle-\langle \overline{E}|x^{m}p^{n-2}x^2|E\rangle-2ig\langle\overline{E}| x^{m}p^{n-2}x^3|E\rangle
\end{align}
We can see from \eqref{rr3} that the higher pure $x$ expectations are expressed in terms of $E$, $\langle\overline{E}| x|E\rangle$ and $\langle\overline{E}| x^{2}|E\rangle$. However, we show below that $\langle\overline{E}| x|E\rangle$ and $\langle\overline{E}| x^{2}|E\rangle$ are not independent data. To show this, we will derive another set of recursion relations by considering the test operator $O=p^t$ in \eqref{fund1nh}.
Using  $[x^3,p^t]=-it(t-1)(t-2)p^{t-3}-3t(t-1)p^{t-2}x+3itp^{t-1}x^2$ and other  basic commutation relations such as the ones given above \eqref{r1nh} we get
\begin{align}
\label{r1nhix3}
&t(t-1)\langle\overline{E}|p^{t-2}|E\rangle-2i\,t\langle\overline{E}|p^{t-1}x|E\rangle\nonumber\\[5pt]
&\hspace{0.1cm}+2i\,g\left(i\,t(t-1)(t-2)\langle\overline{E}| p^{t-3}|E\rangle+3t(t-1)\langle\overline{E}|p^{t-2}x|E\rangle-3i\,t\langle\overline{E}|p^{t-1}x^2|E\rangle\right)=0
\end{align}
When $t=1$ the above recursion takes the simple form 
\begin{align}
\label{r1nhix3t1}
\langle\overline{E}|x|E\rangle+3ig\langle\overline{E}|x^2|E\rangle=0
\end{align}
which shows that $\langle\overline{E}|x^2|E\rangle$ is given in terms of $\langle\overline{E}|x|E\rangle$ as claimed above. Once we have the pure $x$ expectations, we get  mixed expectations from \eqref{rr32}.

The equivalent of \eqref{r1} for $t=0$ in the non-Hermitian PT symmetric case takes the form
\begin{align}
\langle\overline{E}|p|E\rangle=0
\end{align}
More generally, setting $m=0$ and $n=2k+1$ where $k\in Z_{\ge 1}$ in \eqref{rr32} we obtain
\begin{align}
\langle\overline{E}|p^n|E\rangle=0,\quad\text{whenever}\,\,n\,\,\text{is\,\,odd}\,.
\end{align}
%
The non-zero pure $p$ expectations are obtained by setting $m=0$ and $n=2k$ where $k\in Z_{\ge 1}$ in \eqref{rr32}. 

With a full control on the various expectations that come up, we can now perform the null bootstrap. 
\subsubsection{$O(g)$}
We set $K_i=3$ in our ansatz \eqref{ansatzladder}. Solving \eqref{SE2nh} we get the following recursion relation for the energy eigenvalues
\begin{align}
E_{n}^{(1)}=E_{n-1}^{(1)} 
\end{align}
On applying the null state condition at $O(g)$ we get the ground state energy as 
\begin{align}
  E_{0}^{(1)}=0 
\end{align}
Combining the above two equations we get the first order correction to the energy eigenvalue
\begin{align}
   E_{n}^{(1)}=0 
\end{align}
Once again, using the solution of \eqref{SE2nh} and imposing energy independence and normalisation as in \eqref{normalisationnh} we get the following for the first order correction to the lowering operator
%
\begin{align}
L_{-}^{(1)}=-\frac{i}{\sqrt{2}}+  \sqrt{2} ip^2 +\sqrt{2} xp +\frac{ix^2}{\sqrt{2}}
\end{align}
For the raising operator we get 
\begin{align}
L_{+}^{(1)}=\frac{i}{\sqrt{2}}+  \sqrt{2} i p^2 -\sqrt{2} xp +\frac{ix^2}{\sqrt{2}}
\end{align}
Unlike in the Hermitian sextic theory, here $L_{+}^{(1)}$ and $L_{-}^{(1)}$ are not hermitian conjugates of each other. 
\subsubsection{$O(g^2)$}
Repeating the same procedure at $O(g^2)$, where we set $K_i=5$ in our ansatz for the lowering operator \eqref{ansatzladder} we get after solving \eqref{SE2nh}
\begin{align}
\label{recc}
    E_{n-1}^{(2)}= \frac{1}{4} \left(15 - 30 {E_{n}}^{(0)} + 4 {E_{n}}^{(2)}\right)
\end{align}
Applying the null state condition  \eqref{NSCnh} at $O(g^2)$, we get
\begin{align}
  E_{0}^{(2)}= \frac{11}{8}  
\end{align}
Solving the energy recursion relation with the ground state energy as above we get
\begin{align}
    E_{n}^{(2)}=\frac 18\left(30n^2+30n+11\right)
\end{align}
which exactly matches the eigenspectrum of the equivalent Hamiltonian  \eqref{eqHigx3} as obtained in  \eqref{eqHev}. The second order correction to the ladder operators after solving the bootstrap equations take the form
\begin{align}
L_{\pm}^{(2)}=-\frac{61 ip}{16\sqrt{2}}\pm\frac{59x}{16\sqrt{2}}\mp\frac{ip^3}{16\sqrt{2}} +\frac{61xp^2}{16\sqrt{2}}\pm\frac{59ix^2p}{16\sqrt{2}}-\frac{23x^3}{16\sqrt{2}}
\end{align}


\section*{Acknowledgments}
We thank Yongwei Guo and Wenliang Li for  clarifying  doubts.
\appendix
%
\section{Traditional Perturbation Theory}
\label{app-TPTsection}
In this Appendix, we derive the corrections to the energy levels and the ladder operators for the sextic anharmonic oscillator using traditional perturbation theory. For corrections to the ladder operator we follow the discussion in \cite{Guo:2023gfi}. 

The lowering and raising operators in the unperturbed theory are defined as 
\begin{align}
\label{App-aadaggerdef}
a\equiv\frac{1}{\sqrt{2}}(x+ip),\quad a^\dagger\equiv\frac{1}{\sqrt{2}}(x-ip)
\end{align}
and their action on the eigenstates of the unperturbed Hamiltonian is 
\begin{align}
\label{aadaggerdef}
a|E_n\rangle^{(0)}=\sqrt{n}|E_{n-1}\rangle^{(0)},\quad a^\dagger|E_n\rangle^{(0)}=\sqrt{n+1}|E_{n+1}\rangle^{(0)}
\end{align}
In terms of these operators, the perturbation takes the form  
\begin{align}
\label{pert-aadagger}
H'=x^6=\frac 18(a+a^\dagger)^6
\end{align}
First order correction to the energy eigenvalues is given by
\begin{align}
\label{en1app}
E_n^{(1)}&=\prescript{(0)}{}\langle E_n^|H'|E_n\rangle^{(0)}\cr
&=\frac{15}{8}+5n+ \frac{15 n^{2}}{4} + \frac{5 n^{3}}{2}
\end{align}
Second order correction to the energy eigenvalues is given by
\begin{align}
\label{en2app}
E_n^{(2)}&=\sum_{m\ne n}\frac{|\prescript{(0)}{}\langle E_m|H'|E_n\rangle^{(0)}|^2}{E_n^{(0)}-E_m^{(0)}}\cr
&=\frac{1}{64}\left(-3495- 11528 n- 14400 n^2- 12220 n^3-3930 n^4- 1572 n^5\right)
\end{align}
These precisely match the expressions we obtained in \eqref{En1B} and \eqref{En2B}, respectively.

First order correction to the energy eigenstates is given by
\begin{align}
|E_n\rangle^{(1)}=\sum_{m\ne n}\prescript{(0)}{}\langle E_m|E_n\rangle^{(1)}|E_m\rangle^{(0)}+\text{arbitrary\,coefficient}|E_n\rangle^{(0)}
\end{align}
where the coefficients are given by
\begin{align}
\prescript{(0)}{}\langle E_m|E_n\rangle^{(1)}=\frac{\prescript{(0)}{}\langle E_m|H'|E_n\rangle^{(0)}}{(E_n^{(0)}-E_m^{(0)})}
\end{align}
The arbitrariness in the coefficient of the component along $|E_n\rangle^{(0)}$ is fixed by the scheme that the norm of the state, $|E_n\rangle$,  is independent of $g$ up to $O(g)$, i.e. we impose
\begin{align}
\langle E_n|E_n\rangle&=\prescript{(0)}{}\langle E_n|E_n\rangle^{(0)}+g\left(\prescript{(1)}{}\langle E_n|E_n\rangle^{(0)}+\prescript{(0)}{}\langle E_n|E_n\rangle^{(1)}\right)+O(g^2)\cr
&=1+O(g^2)
\end{align}
This leads to 
\begin{align}
\label{En1TPT}
|E_n\rangle^{(1)}&=\frac{1}{96}\left(270a^2+45a^4+2 a^6- 270 {a^\dag}^2 -45 {a^\dag}^4-2{a^\dag}^6 + 
   360 {a^\dag} a^3 + 18 a^\dag a^5\right.\cr
  & \hspace{1.5cm}\left.+90 {{a^\dag}^2}a^4-360 {{a^\dag}^3}a - 
   90{{a^\dag}^4}a^2- 18 {{a^\dag}^5}a\right)|E_n\rangle^{(0)}\cr
&\equiv f^{(1)}|E_n\rangle^{(0)}
\end{align}
Second order correction to the energy eigenstates is given by
\begin{align}
\label{en2exp}
|E_n\rangle^{(2)}=\sum_{m\ne n}\prescript{(0)}{}\langle E_m|E_n\rangle^{(2)}|E_m\rangle^{(0)}+\text{arbitrary\,coefficient}|E_n\rangle^{(0)}
\end{align}
where the coefficients are 
\begin{align}
\prescript{(0)}{}\langle E_m|E_n\rangle^{(2)}=\frac{\prescript{(0)}{}\langle E_m|\left(H'-E_n^{(1)}\right)f^{(1)}|E_n\rangle^{(0)}}{E_n^{(0)}-E_m^{(0)}}
\end{align}
Once again, the coefficient of $|E_n\rangle^{(0)}$ in the expansion \eqref{en2exp} remains arbitrary and we fix it by requiring that the norm of the state is independent of $g$ up to $O(g^2)$, i.e. we impose
\begin{align}
\langle E_n|E_n\rangle&=\prescript{(0)}{}\langle E_n|E_n\rangle^{(0)}+g\left(\prescript{(1)}{}\langle E_n|E_n\rangle^{(0)}+\prescript{(0)}{}\langle E_n|E_n\rangle^{(1)}\right)\cr
&\hspace{0.5cm}+g^2\left(\prescript{(2)}{}\langle E_n|E_n\rangle^{(0)}+\prescript{(0)}{}\langle E_n|E_n\rangle^{(2)}+\prescript{(1)}{}\langle E_n|E_n\rangle^{(1)}\right)\cr
&=1+O(g^3)
\end{align}
This gives 
\begin{align}
\label{En2TPT}
|E_n\rangle^{(2)}&=\left(\frac{-1755 a^2}{8} - \frac{19575 a^4}{512} + \frac{11405 a^6}{1536} + \frac{
 2325 a^8}{2048} + \frac{61 a^{10}}{2560} + \frac{a^{12}}{4608} + \frac{18495 {a^\dag}^2}{128} \right. \cr
 &\hspace{1cm}+ \frac{
 27825 {a^\dag}^4}{512} + \frac{13165 {a^\dag}^6}{1536} + \frac{1245 {a^\dag}^8}{2048}+ \frac{
 49 {a^\dag}^{10}}{2560} + \frac{{a^\dag}^{12}}{4608} - \frac{22275 {a^\dag} a^3}{32}\cr
 &\hspace{1cm} - \frac{
 1215 {a^\dag } a^5}{64} + \frac{2245 {a^\dag}a^7}{256} + \frac{55 {a^\dag}a^9}{128} + 
 \frac{{a^\dag} a^{11}}{256} - \frac{300375 {{a^\dag}^2} a^4}{512} + \frac{165{a^\dag}^2  a^6}{16} \cr
&\hspace{1cm} + \frac{
 1225 {{a^\dag}^2}a^8}{512} + \frac{19 {{a^\dag}^2} a^{10}}{512 }+ \frac{15825 {{a^\dag}^3} a}{32} - \frac{
 42885{{a^\dag}^3} a^5}{256} + \frac{155{{a^\dag}^3 } a^7}{32} \cr
&\hspace{1cm} + \frac{45 {{a^\dag}^3}a^9}{256} + \frac{222825 {{a^\dag}^4} a^2}{512} - \frac{3895 {{a^\dag}^4}a^6}{256} + \frac{
 215 {{a^\dag}^4} a^8}{512} + \frac{6195 {{a^\dag}^5}a}{64} \cr
  &\hspace{1cm}+ \frac{31275{{a^\dag}^5} a^3}{256}-\frac{23{{a^\dag}^5} a^7}{128} + \frac{795{{a^\dag}^6} a^2}{16} 
 + \frac{2285 {{a^\dag}^6}a^4}{256} + \frac{
 2165 {{a^\dag}^7} a}{256}\cr&\hspace{1cm} + \frac{275{{a^\dag}^7} a^3}{32} 
 - \frac{23 {{a^\dag}^7}a^5}{128}+\frac{1205{{a^\dag}^8} a^2}{512} + \frac{215{{a^\dag}^8} a^4}{512} + \frac{5{{a^\dag}^9 }a}{16} + \frac{
 45 {{a^\dag}^9} a^3}{256}\cr&\hspace{1cm} + \frac{19{{a^\dag}^{10}}a^2}{512} + \frac{{{a^\dag}^{11}}a}{256}
 -\frac{18615{a^\dag}a}{128}-\frac{46725{{a^\dag}^2}a^2}{128}-\frac{116975{{a^\dag}^3}a^3}{384}\cr
&\hspace{1cm}\left. -\frac{109225{{a^\dag}^4}a^4}{1024}-\frac{2107{{a^\dag}^5}a^5}{128}-\frac{2107{{a^\dag}^6}a^6}{2304}-\frac{685}{64}\right)|E_n\rangle^{(0)}\cr
&\equiv f^{(2)}|E_n\rangle^{(0)}
\end{align}
Let us now obtain the corrections to the ladder operator at different orders in the coupling. Our starting equation is the defining equation of ladder operators as below
\begin{align}
\label{app-ladder-def}
L_{\pm}|E_n\rangle=C_{\pm}|E_{n\pm 1}\rangle
\end{align}
where $C_{\pm}$ are given by
\begin{align}
C_{+}&=\sqrt{n+1}\cr
C_{-}&=\sqrt{n}
\end{align}
This choice of $C_{\pm}$ is consistent with the normalisation in \eqref{normalisation}.
Plugging the series expansion of the lowering operator \eqref{ladderseriesassumption} and that of the energy eigenstate in \eqref{app-ladder-def} we obtain at the first few orders the following 
%
\begin{align}
L_{-}^{(0)}|E_n\rangle^{(0)}=\sqrt{n}|E_{n-1}\rangle^{(0)}\cr
L_{-}^{(0)}|E_n\rangle^{(1)}+L_{-}^{(1)}|E_n\rangle^{(0)}=\sqrt{n}|E_{n-1}\rangle^{(1)}\cr
L_{-}^{(0)}|E_n\rangle^{(2)}+L_{-}^{(1)}|E_n\rangle^{(1)}+L_{-}^{(2)}|E_n\rangle^{(0)}=\sqrt{n}|E_{n-1}\rangle^{(2)}
\end{align}
The first of these corresponds to the definition of the lowering operator in the unperturbed theory \eqref{aadaggerdef}. We have
\begin{align}
\label{Lminus0}
L_{-}^{(0)}=a=\frac{1}{\sqrt{2}}(x+ip)
\end{align}
In the second and third equations, we make use of the definition of $f_1$ and $f_2$ from \eqref{En1TPT} and \eqref{En2TPT} respectively and obtain
\begin{align}
L_{-}^{(0)}f^{(1)}|E_n\rangle^{(0)}+L_{-}^{(1)}|E_n\rangle^{(0)}&=f^{(1)}L_{-}^{(0)}|E_n\rangle^{(0)}\cr
L_{-}^{(0)}f^{(2)}|E_n\rangle^{(0)}+L_{-}^{(1)}f^{(1)}|E_n\rangle^{(0)}+L_{-}^{(2)}|E_n\rangle^{(0)}&=f^{(2)}L_{-}^{(0)}|E_n\rangle^{(0)}
\end{align}
From these equations we obtain 
  \begin{align}
\label{eq(B.1)}
     L_{-}^{(1)}&=[f^{(1)},L_{-}^{(0)}]\cr
    L_{-}^{(2)}&=[f^{(2)},L_{-}^{(0)}]-L_{-}^{(1)}f^{(1)}
    \end{align}
    A similar derivation gives the following for the raising operator
      \begin{align}
\label{eq(B.6)}
L_{+}^{(0)}&=a^\dagger=\frac{1}{\sqrt{2}}(x-ip)\cr
     L_{+}^{(1)}&=[f^{(1)},L_{+}^{(0)}]\cr
    L_{+}^{(2)}&=[f^{(2)},L_{+}^{(0)}]-L_{+}^{(1)}f^{(1)}
    \end{align}
  We now make use of the expressions for $f_1$ and $f_2$ from \eqref{En1TPT} and \eqref{En2TPT} respectively  and $L_{-}^{(0)}$ from \eqref{Lminus0} and obtain
  \begin{align}\label{laddappexp}
  L_{-}^{(1)}&=\frac{90 (a^{\dag})}{16}-\frac{15 a^{3}}{4}+\frac{45(a^{\dag})^{2}a}{4}+\frac{60(a^{\dag})^{3}}{32}-\frac{3a^{5}}{16}-\frac{30a^{\dag}a^4}{16}+\frac{60(a^{\dag})^{3}a^{2}}{16}+\frac{15(a^{\dag})^{4}a}{16}\cr
&\hspace{.5cm}+\frac{6(a^{\dag})^{5}}{48}\cr
    L_{-}^{(2)}&=\frac{1}{512}\left[-8700a+318600a^{3}+47520a^{5}+1360a^{7}-45a^{9}-181800a^{\dag}\right.\cr
 & \hspace{.5cm}\left.-72000(a^{\dag})^{3} +2020(a^{\dag})^{5}+1020(a^{\dag})^{7}+12(a^{\dag})^{9}-1500a^{\dag}a^{2}+553500a^{\dag}a^{4}\right.\cr
&\hspace{.5cm}\left. +35280(a^{\dag})a^{6}+340(a^{\dag})a^{8}-882000(a^{\dag})^{2}a+30500(a^{\dag})^{2}a^{3}+236520(a^{\dag})^{2}a^{5}\right.\cr
&\hspace{.5cm}\left.+5040(a^{\dag})^{2}a^{7}-1004400(a^{\dag})^{3}a^{2}+15500(a^{\dag})^{3}a^{4}+26280(a^{\dag})^{3}a^{6}\right.\cr
&\hspace{.5cm}\left.-104400(a^{\dag})^{4}a-355200(a^{\dag})^{4}a^{3}
+1550(a^{\dag})^{4}a^{5}-41040(a^{\dag})^{5}a^{2}
-35520(a^{\dag})^{5}a^{4}\right.\cr
&\hspace{.5cm}\left.+3080(a^{\dag})^{6}a-4560(a^{\dag})^{6}a^{3}+440(a^{\dag})^{7}a^{2}+255(a^{\dag})^{8}a\right]
\end{align}
Similarly, we make use of \eqref{eq(B.6)} and get
\begin{align}
\label{laddappexp1}
 L_{+}^{(1)}&={L_{-}^{(1)}}^\dagger\cr
  L_{+}^{(2)}&={L_{-}^{(2)}}^\dagger
\end{align}
where ${L_{-}^{(1)}}^\dagger$ and ${L_{-}^{(2)}}^\dagger$ denote the Hermitian conjugates of ${L_{-}^{(1)}}$ and ${L_{-}^{(2)}}$ in \eqref{laddappexp}.

 \bibliography{ref.bib}

\providecommand{\href}[2]{#2}\begingroup\raggedright\begin{thebibliography}{10}

\bibitem{Poland:2018epd}
D.~Poland, S.~Rychkov and A.~Vichi, \emph{{The Conformal Bootstrap: Theory,
  Numerical Techniques, and Applications}},
  \href{https://doi.org/10.1103/RevModPhys.91.015002}{\emph{Rev. Mod. Phys.}
  {\bfseries 91} (2019) 015002}
  [\href{https://arxiv.org/abs/1805.04405}{{\ttfamily 1805.04405}}].

\bibitem{Lin:2020mme}
H.~W. Lin, \emph{{Bootstraps to strings: solving random matrix models with
  positivity}}, \href{https://doi.org/10.1007/JHEP06(2020)090}{\emph{JHEP}
  {\bfseries 06} (2020) 090}
  [\href{https://arxiv.org/abs/2002.08387}{{\ttfamily 2002.08387}}].

\bibitem{Han:2020bkb}
X.~Han, S.~A. Hartnoll and J.~Kruthoff, \emph{{Bootstrapping Matrix Quantum
  Mechanics}},
  \href{https://doi.org/10.1103/PhysRevLett.125.041601}{\emph{Phys. Rev. Lett.}
  {\bfseries 125} (2020) 041601}
  [\href{https://arxiv.org/abs/2004.10212}{{\ttfamily 2004.10212}}].

\bibitem{Kazakov:2021lel}
V.~Kazakov and Z.~Zheng, \emph{{Analytic and numerical bootstrap for one-matrix
  model and \textquotedblleft{}unsolvable\textquotedblright{} two-matrix
  model}}, \href{https://doi.org/10.1007/JHEP06(2022)030}{\emph{JHEP}
  {\bfseries 06} (2022) 030}
  [\href{https://arxiv.org/abs/2108.04830}{{\ttfamily 2108.04830}}].

\bibitem{Berenstein:2021dyf}
D.~Berenstein and G.~Hulsey, \emph{{Bootstrapping Simple QM Systems}},
  \href{https://arxiv.org/abs/2108.08757}{{\ttfamily 2108.08757}}.

\bibitem{Bhattacharya:2021btd}
J.~Bhattacharya, D.~Das, S.~K. Das, A.~K. Jha and M.~Kundu, \emph{{Numerical
  bootstrap in quantum mechanics}},
  \href{https://doi.org/10.1016/j.physletb.2021.136785}{\emph{Phys. Lett. B}
  {\bfseries 823} (2021) 136785}
  [\href{https://arxiv.org/abs/2108.11416}{{\ttfamily 2108.11416}}].

\bibitem{Aikawa:2021eai}
Y.~Aikawa, T.~Morita and K.~Yoshimura, \emph{{Application of bootstrap to a
  \ensuremath{\theta} term}},
  \href{https://doi.org/10.1103/PhysRevD.105.085017}{\emph{Phys. Rev. D}
  {\bfseries 105} (2022) 085017}
  [\href{https://arxiv.org/abs/2109.02701}{{\ttfamily 2109.02701}}].

\bibitem{Berenstein:2021loy}
D.~Berenstein and G.~Hulsey, \emph{{Bootstrapping more QM systems}},
  \href{https://doi.org/10.1088/1751-8121/ac7118}{\emph{J. Phys. A} {\bfseries
  55} (2022) 275304} [\href{https://arxiv.org/abs/2109.06251}{{\ttfamily
  2109.06251}}].

\bibitem{Tchoumakov:2021mnh}
S.~Tchoumakov and S.~Florens, \emph{{Bootstrapping Bloch bands}},
  \href{https://doi.org/10.1088/1751-8121/ac3c82}{\emph{J. Phys. A} {\bfseries
  55} (2022) 015203} [\href{https://arxiv.org/abs/2109.06600}{{\ttfamily
  2109.06600}}].

\bibitem{Aikawa:2021qbl}
Y.~Aikawa, T.~Morita and K.~Yoshimura, \emph{{Bootstrap method in harmonic
  oscillator}},
  \href{https://doi.org/10.1016/j.physletb.2022.137305}{\emph{Phys. Lett. B}
  {\bfseries 833} (2022) 137305}
  [\href{https://arxiv.org/abs/2109.08033}{{\ttfamily 2109.08033}}].

\bibitem{Bai:2022yfv}
D.~Bai, \emph{{Bootstrapping the deuteron}},
  \href{https://arxiv.org/abs/2201.00551}{{\ttfamily 2201.00551}}.

\bibitem{Nakayama:2022ahr}
Y.~Nakayama, \emph{{Bootstrapping microcanonical ensemble in classical
  system}}, \href{https://doi.org/10.1142/S0217732322500547}{\emph{Mod. Phys.
  Lett. A} {\bfseries 37} (2022) 2250054}
  [\href{https://arxiv.org/abs/2201.04316}{{\ttfamily 2201.04316}}].

\bibitem{Li:2022prn}
W.~Li, \emph{{Null bootstrap for non-Hermitian Hamiltonians}},
  \href{https://doi.org/10.1103/PhysRevD.106.125021}{\emph{Phys. Rev. D}
  {\bfseries 106} (2022) 125021}
  [\href{https://arxiv.org/abs/2202.04334}{{\ttfamily 2202.04334}}].

\bibitem{Hu:2022keu}
X.~Hu, \emph{{Different Bootstrap Matrices in Many QM Systems}},
  \href{https://arxiv.org/abs/2206.00767}{{\ttfamily 2206.00767}}.

\bibitem{Blacker:2022szo}
M.~J. Blacker, A.~Bhattacharyya and A.~Banerjee, \emph{{Bootstrapping the
  Kronig-Penney model}},
  \href{https://doi.org/10.1103/PhysRevD.106.116008}{\emph{Phys. Rev. D}
  {\bfseries 106} (2022) 116008}
  [\href{https://arxiv.org/abs/2209.09919}{{\ttfamily 2209.09919}}].

\bibitem{Berenstein:2022unr}
D.~Berenstein and G.~Hulsey, \emph{{Semidefinite programming algorithm for the
  quantum mechanical bootstrap}},
  \href{https://doi.org/10.1103/PhysRevE.107.L053301}{\emph{Phys. Rev. E}
  {\bfseries 107} (2023) L053301}
  [\href{https://arxiv.org/abs/2209.14332}{{\ttfamily 2209.14332}}].

\bibitem{Fan:2023bld}
W.~Fan and H.~Zhang, \emph{{Non-perturbative instanton effects in the quartic
  and the sextic double-well potential by the numerical bootstrap approach}},
  \href{https://arxiv.org/abs/2308.11516}{{\ttfamily 2308.11516}}.

\bibitem{Guo:2023gfi}
Y.~Guo and W.~Li, \emph{{Solving anharmonic oscillator with null states:
  Hamiltonian bootstrap and Dyson-Schwinger equations}},
  \href{https://arxiv.org/abs/2305.15992}{{\ttfamily 2305.15992}}.

\bibitem{Bender:1998ke}
C.~M. Bender and S.~Boettcher, \emph{{Real spectra in nonHermitian Hamiltonians
  having PT symmetry}},
  \href{https://doi.org/10.1103/PhysRevLett.80.5243}{\emph{Phys. Rev. Lett.}
  {\bfseries 80} (1998) 5243}
  [\href{https://arxiv.org/abs/physics/9712001}{{\ttfamily physics/9712001}}].

\bibitem{Dorey:2001uw}
P.~Dorey, C.~Dunning and R.~Tateo, \emph{{Spectral equivalences, Bethe Ansatz
  equations, and reality properties in PT-symmetric quantum mechanics}},
  \href{https://doi.org/10.1088/0305-4470/34/28/305}{\emph{J. Phys. A}
  {\bfseries 34} (2001) 5679}
  [\href{https://arxiv.org/abs/hep-th/0103051}{{\ttfamily hep-th/0103051}}].

\bibitem{Mostafazadeh:2001jk}
A.~Mostafazadeh, \emph{{PseudoHermiticity versus PT symmetry. The necessary
  condition for the reality of the spectrum}},
  \href{https://doi.org/10.1063/1.1418246}{\emph{J. Math. Phys.} {\bfseries 43}
  (2002) 205} [\href{https://arxiv.org/abs/math-ph/0107001}{{\ttfamily
  math-ph/0107001}}].

\bibitem{Mostafazadeh:2001nr}
A.~Mostafazadeh, \emph{{PseudoHermiticity versus PT symmetry 2. A Complete
  characterization of nonHermitian Hamiltonians with a real spectrum}},
  \href{https://doi.org/10.1063/1.1461427}{\emph{J. Math. Phys.} {\bfseries 43}
  (2002) 2814} [\href{https://arxiv.org/abs/math-ph/0110016}{{\ttfamily
  math-ph/0110016}}].

\bibitem{Mostafazadeh:2002id}
A.~Mostafazadeh, \emph{{PseudoHermiticity versus PT symmetry 3: Equivalence of
  pseudoHermiticity and the presence of antilinear symmetries}},
  \href{https://doi.org/10.1063/1.1489072}{\emph{J. Math. Phys.} {\bfseries 43}
  (2002) 3944} [\href{https://arxiv.org/abs/math-ph/0203005}{{\ttfamily
  math-ph/0203005}}].

\bibitem{Solombrino:2002vk}
L.~Solombrino, \emph{{Weak pseudo-Hermiticity and antilinear commutant}},
  \href{https://doi.org/10.1063/1.1504485}{\emph{J. Math. Phys.} {\bfseries 43}
  (2002) 5439} [\href{https://arxiv.org/abs/quant-ph/0203101}{{\ttfamily
  quant-ph/0203101}}].

\bibitem{Bender:2002vv}
C.~M. Bender, D.~C. Brody and H.~F. Jones, \emph{{Complex extension of quantum
  mechanics}}, \href{https://doi.org/10.1103/PhysRevLett.89.270401}{\emph{Phys.
  Rev. Lett.} {\bfseries 89} (2002) 270401}
  [\href{https://arxiv.org/abs/quant-ph/0208076}{{\ttfamily
  quant-ph/0208076}}].

\bibitem{Mostafazadeh:2002pd}
A.~Mostafazadeh, \emph{{PseudoHermiticity and generalized PT and CPT
  symmetries}}, \href{https://doi.org/10.1063/1.1539304}{\emph{J. Math. Phys.}
  {\bfseries 44} (2003) 974}
  [\href{https://arxiv.org/abs/math-ph/0209018}{{\ttfamily math-ph/0209018}}].

\bibitem{Bender:2003fi}
C.~M. Bender, P.~N. Meisinger and Q.~Wang, \emph{{Calculation of the hidden
  symmetry operator in PT-symmetric quantum mechanics}},
  \href{https://doi.org/10.1088/0305-4470/36/7/312}{\emph{J. Phys. A}
  {\bfseries 36} (2003) 1973}
  [\href{https://arxiv.org/abs/quant-ph/0211166}{{\ttfamily
  quant-ph/0211166}}].

\bibitem{Mostafazadeh:2003gz}
A.~Mostafazadeh, \emph{{Exact PT symmetry is equivalent to Hermiticity}},
  \href{https://doi.org/10.1088/0305-4470/36/25/312}{\emph{J. Phys. A}
  {\bfseries 36} (2003) 7081}
  [\href{https://arxiv.org/abs/quant-ph/0304080}{{\ttfamily
  quant-ph/0304080}}].

\bibitem{Bender:2004sa}
C.~M. Bender, D.~C. Brody and H.~F. Jones, \emph{{Extension of PT symmetric
  quantum mechanics to quantum field theory with cubic interaction}},
  \href{https://doi.org/10.1103/PhysRevD.70.025001}{\emph{Phys. Rev. D}
  {\bfseries 70} (2004) 025001}
  [\href{https://arxiv.org/abs/hep-th/0402183}{{\ttfamily hep-th/0402183}}].

\bibitem{Bender:2004vn}
C.~M. Bender, D.~C. Brody and H.~F. Jones, \emph{{Scalar quantum field theory
  with cubic interaction}},
  \href{https://doi.org/10.1103/PhysRevLett.93.251601}{\emph{Phys. Rev. Lett.}
  {\bfseries 93} (2004) 251601}
  [\href{https://arxiv.org/abs/hep-th/0402011}{{\ttfamily hep-th/0402011}}].

\bibitem{Mostafazadeh:2004qh}
A.~Mostafazadeh, \emph{{PT-symmetric cubic anharmonic oscillator as a physical
  model}}, \href{https://doi.org/10.1088/0305-4470/38/29/010}{\emph{J. Phys. A}
  {\bfseries 38} (2005) 6557}
  [\href{https://arxiv.org/abs/quant-ph/0411137}{{\ttfamily
  quant-ph/0411137}}].

\bibitem{Mostafazadeh:2005wm}
A.~Mostafazadeh, \emph{{Metric operator in pseudo-Hermitian quantum mechanics
  and the imaginary cubic potential}},
  \href{https://doi.org/10.1088/0305-4470/39/32/S18}{\emph{J. Phys. A}
  {\bfseries 39} (2006) 10171}
  [\href{https://arxiv.org/abs/quant-ph/0508195}{{\ttfamily
  quant-ph/0508195}}].

\bibitem{Bender:2005tb}
C.~M. Bender, \emph{{Introduction to PT-Symmetric Quantum Theory}},
  \href{https://doi.org/10.1080/00107500072632}{\emph{Contemp. Phys.}
  {\bfseries 46} (2005) 277}
  [\href{https://arxiv.org/abs/quant-ph/0501052}{{\ttfamily
  quant-ph/0501052}}].

\bibitem{Bender:2006xa}
C.~M. Bender and B.~Tan, \emph{{Calculation of the Hidden Symmetry Operator for
  a PT-Symmetric Square Well}},
  \href{https://doi.org/10.1088/0305-4470/39/8/011}{\emph{J. Phys. A}
  {\bfseries 39} (2006) 1945}
  [\href{https://arxiv.org/abs/quant-ph/0601123}{{\ttfamily
  quant-ph/0601123}}].

\bibitem{Bender:2007nj}
C.~M. Bender, \emph{{Making sense of non-Hermitian Hamiltonians}},
  \href{https://doi.org/10.1088/0034-4885/70/6/R03}{\emph{Rept. Prog. Phys.}
  {\bfseries 70} (2007) 947}
  [\href{https://arxiv.org/abs/hep-th/0703096}{{\ttfamily hep-th/0703096}}].

\bibitem{Bender:2007wb}
C.~M. Bender and H.~F. Jones, \emph{{Interactions of Hermitian and
  non-Hermitian Hamiltonians}},
  \href{https://doi.org/10.1088/1751-8113/41/24/244006}{\emph{J. Phys. A}
  {\bfseries 41} (2008) 244006}
  [\href{https://arxiv.org/abs/0709.3605}{{\ttfamily 0709.3605}}].

\bibitem{Mostafazadeh:2008pw}
A.~Mostafazadeh, \emph{{Pseudo-Hermitian Representation of Quantum Mechanics}},
  \href{https://doi.org/10.1142/S0219887810004816}{\emph{Int. J. Geom. Meth.
  Mod. Phys.} {\bfseries 7} (2010) 1191}
  [\href{https://arxiv.org/abs/0810.5643}{{\ttfamily 0810.5643}}].

\bibitem{Mannheim:2017apd}
P.~D. Mannheim, \emph{{Appropriate Inner Product for PT-Symmetric
  Hamiltonians}}, \href{https://doi.org/10.1103/PhysRevD.97.045001}{\emph{Phys.
  Rev. D} {\bfseries 97} (2018) 045001}
  [\href{https://arxiv.org/abs/1708.01247}{{\ttfamily 1708.01247}}].

\bibitem{Khan:2022uyz}
S.~Khan, Y.~Agarwal, D.~Tripathy and S.~Jain, \emph{{Bootstrapping PT symmetric
  quantum mechanics}},
  \href{https://doi.org/10.1016/j.physletb.2022.137445}{\emph{Phys. Lett. B}
  {\bfseries 834} (2022) 137445}
  [\href{https://arxiv.org/abs/2202.05351}{{\ttfamily 2202.05351}}].

\end{thebibliography}\endgroup
\bibliographystyle{JHEP}


\end{document}